\documentclass[reprint,aip,pop]{revtex4-1}
\usepackage{graphicx}

\begin{document}
\title{Evaluation of Charged Particle Evaporation Expressions in Ultracold Plasmas}
\author{Craig Witte and Jacob L. Roberts} 
\address{Colorado State University, Fort Collins CO, 80523}
\begin{abstract}
Electron evaporation plays an important role in the electron temperature evolution and thus expansion rate in low-density ultracold plasmas. In addition, evaporation is useful as a potential tool for obtaining colder electron temperatures and characterizing plasma parameters. Evaporation theory has been developed for atomic gases and has been applied to a one-component plasma system. We numerically investigate whether such an adapted theory is applicable to ultracold neutral plasmas. We find that it is not due to the violation of fundamental assumptions of the model. The details of our calculations are presented as well as a discussion of the implications for a simple description of the electron evaporation rate in ultracold plasmas.
\end{abstract}
\maketitle

\section{Introduction}

Ultracold plasmas (UCPs) offer the opportunity to study plasma physics within a unique range of plasma parameters.\cite{killian1999,physicstoday} Their low temperatures and controllable initial conditions make these plasmas good candidates to explore fundamental plasma physics as well as strong coupling physics. Electron temperatures play a critical role in establishing electron equilibration times, screening, and strong coupling effects\cite{strong_coupling_review} in the electron component of the UCP. There are many influences on the electron temperature in UCPs, including three-body recombination\cite{killian2001}, continuum lowering heating\cite{sp}, disorder-induced heating\cite{Lyon2011}, cooling via UCP expansion\cite{simple}, and evaporative cooling\cite{Wilson2013}. Electron evaporation occurs when electrons escape from the UCP's confining potential and leave the plasma. Since only high energy electrons are able to escape from the UCP, evaporation leads to a net energy loss from the trapped electron cloud that results in a lower electron temperature. Electron evaporation is especially important at low densities, where a large fraction of electrons are able to escape\cite{Wilson2013}. 

Electron evaporation, when properly understood, should offer a variety of insights into some of the fundamental plasma physics associated with UCPs. Evaporation can have a sizable impact on the expansion dynamics of UCPs as electron evaporation results in a cooling of the electron component, reducing the rate of expansion\cite{simple}. Furthermore, electron evaporation leads to higher levels of charge imbalance, resulting in additional Coulomb forces that distort the plasma expansion\cite{Witte14}. Additionally, strong coupling physics can be accessed in UCPs due to their low temperatures. Evaporation-induced cooling of the electron component should allow access to a greater degree of strong coupling.

Finally, electron evaporation is a good candidate to probe the temperature of the electron cloud. Techniques for measuring UCP ion temperatures are well established\cite{killian2005}, but such techniques are not applicable to UCP electrons, and typically electron temperatures are estimated based on theoretical interpretations of ion expansion\cite{Gupta2007} or derived from other plasma properties\cite{Roberts2004,Fletcher2007}. In the absence of other effects, electron temperatures would be solely determined by the photon energy of the ionizing laser pulse. However, such a simple estimation ignores various heating mechanisms such as continuum lowering, disorder induced heating, and three-body recombination\cite{killian2001,sp,Lyon2011}. Theoretical predictions quantifying these heating mechanisms do exist, and can be incorporated into temperature estimates, but tests of the predicted heating rates are not available across all UCP parameter ranges. Furthermore, it is not clear how well these heating predictions extend into the strongly coupled regime.

The development of an independent electron temperature measurement would offer the ability to test the net temperature change due to the previously listed effects. UCP electron evaporation is a good candidate for such measurement over a wide range of UCP conditions, since there is a strong electron temperature dependence expected in the electron evaporation rate, making the measurement of the rate potentially sensitive to the electron temperature.

For the electron evaporation rate to be used to determine the electron temperature, the electron evaporation rate and the electron temperature needs to be linked. It would be ideal if analytical expressions describing the functional form of the electron evaporation rate as a function of electron temperature were known. In the context of ultracold atomic gases, such expressions have long been established\cite{coolingrev}. More recently, the ALPHA collaboration has adapted these expressions to be applicable to the antiproton plasmas present in their experimental system\cite{antihydrogen}. Naively, we would expect that such expressions would also be applicable to UCP systems. However, these expressions make two assumptions that are not clearly applicable to UCPs. First, it is implicitly assumed that the average electron mean free path is much greater than the spatial size of the UCP. Second, this treatment ignores the possibility that larger angle Coulomb collisions have a significant role to play in evaporation.

Evaluation of the validity of these assumptions is one of the main topics of this work.  We find that neither assumption is valid under a typical set of UCP conditions.  We can characterize the degree to which these assumptions are broken.  Our results indicate that a simple analytic description of UCP electron evaporation rate would be highly challenging to formulate. The other main part of this work is the model that we have developed, which allows for making quantitative predictions of electron evaporation rates and can be directly applicable to experimentally-relevant parameters with straightforward modification.

In Sec II, we discuss the theoretical treatment of evaporation in ultracold atomic systems, how this treatment has previously been adapted to plasma systems, and potential issues that could arise from applying this adapted treatment to ultracold plasmas. Sec III gives an overview of how our theoretical evaporation model functions. In Sec IV, we report our numerical results, as well as use these results to test whether previous plasma evaporation treatment can be applied to ultracold plasma systems. Finally, in Sec V, we present our conclusions and possibilities for future work. 

\section{Theory}

Evaporation is a commonly used experimental technique in ultracold atomic physics, where collections of atoms are routinely trapped inside potential wells produced by external optical or magnetic fields\cite{coolingrev,Anderson1995,Davis95,Barrett01}. In these systems, evaporation occurs when an atom gains sufficient energy to be able to escape from the trapping potential. The threshold energy necessary for escape is equivalent to the overall depth of the trapping potential, and will be henceforth referred to as the potential barrier, $U$.

While confined, atoms in the cloud collide with each other. These collisions lead to energy being transfered between the atoms and can lead to an atom acquiring energy greater than the barrier. If an atom is assumed to escape once its energy exceeds the barrier, $U$, the evaporation rate would in that case be the rate at which collisions excite atoms above the barrier.

Integrating over all possible elastic collisions in a thermal gas is difficult, making the determination of the evaporation rate a non-trivial endeavor. Traditionally, in neutral atoms, this difficulty has been mitigated by utilizing the principle of detailed balance. Since collisions with neutral atoms are predominantly large angle collisions, it is reasonable to assume that the vast majority of collisions involving an atom with energy exceeding $U$ will cause that atom to lose energy and fall back below the barrier. Detailed balance indicates that rate at which particles fall below barrier is equivalent to the rate at which atoms are excited above the barrier in thermal equilibrium. This leads to the following approximate general expression for the rate of change in particle number from evaporation\cite{coolingrev}:

\begin{equation}
\dot{N}=-n\sigma N_{he}v_{U}=-N_{he}\nu_{col}
\end{equation}
where $n$ is the particle density, $\sigma$ is the collisional cross-section, $N_{he}$ is the number of high energy atoms with energy exceeding the barrier, $v_{U}$ is the atom velocity that corresponds to the barrier energy, and $\nu_{col}$ is the average collision frequency for high energy atoms in the distribution. If the atoms are assumed to be distributed in a Maxwellian with three degrees of freedom, and the barrier energy to average thermal energy ratio is sufficiently high, the fraction of atoms above the barrier is approximately $2\sqrt{W/\pi}e^{-W}$. In this limit Eq. (1) takes the following form\cite{coolingrev}:

\begin{equation}
\dot{N}=-n\sigma We^{-W}\bar{v}N
\end{equation}
where $\bar{v}$ is the average velocity in the particle distribution and $W$ is the scaled barrier height defined as $U/k_bT$.

In plasmas, the presence of charged particles substantially alters the dynamics of evaporation. In general, Coulomb collisions have a much larger interaction range than the hard sphere collisions involved in collisions of neutrals. This leads to a much larger collisional cross-section, which in turn results in a substantially higher frequency of collisions. Furthermore, Coulomb collision cross-sections are velocity dependent, necessitating additional care when calculating $\dot{N}$. Finally, the Coulomb interaction leads predominantly to small angle deflections from collisions\cite{Fund_PP}. This last point has a profound impact with respect to the previous detailed balance assumptions, since it is no longer reasonable to assume that a collision will knock a high energy particle below the barrier. In a recent adaptation of atomic evaporation theory to charged particles, an alternative assumption is made. It was assumed that only a small amount of energy is transfered between colliding particles. In the context of electron evaporation, this allows for the evaporation rate to expressed in the following way\cite{antihydrogenthesis}:

\begin{equation}
\dot{N}=\bigg(\frac{dN}{dt}\bigg)\bigg|_{v=v_U}=\bigg(\frac{dN}{dv}\frac{dv}{dt}\bigg)\bigg|_{v=v_U}
\end{equation}
where $dN/dv$ is simply the electrons' velocity distribution, and $dv/dt$ is a velocity damping rate. The negative sign enters the above equation from detailed balance. For a Maxwell Boltzmann distribution with three degrees of freedom, $\frac{dN}{dv}$ takes the following form\cite{antihydrogenthesis}:

\begin{equation}
\bigg(\frac{dN}{dv}\bigg)=4\pi Nv^2\bigg(\frac{m}{2\pi k_bT}\bigg)^{3/2}e^{\frac{-mv^2}{2k_bT}}
\end{equation}
In the limit of $v$ being much larger than the average thermal velocity, $dv/dt$ can be calculated via conservation of momentum. The result of this calculation is\cite{antihydrogenthesis}:

\begin{equation}
\frac{dv}{dt}=-\frac{2e^4nln(\Lambda)}{4\pi\epsilon_0^2\mu^2v^2}
\end{equation}
where $e$ is the fundamental electron charge, $\mu$ is the reduced mass of the two colliding particles, and $ln(\Lambda)$ is the Coulomb logarithm that results from averaging over all possible Coulomb collision angles in the typical treatment of Coulomb collisions in a plasma. Combining eq3-5 yields the following expression for evaporation\cite{antihydrogenthesis}:

\begin{equation}
\dot{N}=-Ne^{-W}\frac{\sqrt{2}e^4nln(\Lambda)}{\pi^{3/2}\epsilon_0^2\sqrt{\mu}(k_bT)^{3/2}}
\end{equation}

As mentioned above, equation (6) has two underlying assumptions which are not obviously applicable to UCPs. First, it is assumed that once an electron's energy exceeds $U$, it immediately escapes from the plasma. However, it is easy to imagine that once a electron is excited above the barrier subsequent collisions could knock that electron back below the barrier before it is able to escape. Implicitly, immediate escape assumes that the average mean free path for an electron is much larger than the spatial size of the UCP. A naive estimate of the mean free path, $c$, would simply be $v^2/\frac{dv}{dt}$, the velocity times the effective velocity damping time constant. Such an estimate would yield a mean free path for typical experimental conditions roughly  an order of magnitude larger than typical UCP sizes, apparently satisfying the prior assumption.

However, such an estimate implies an average electron slowing of $1/e$ is the relevant amount of velocity decrease for determining an effective mean free path electrons with kinetic energy just greater than $U$. Considering that the overwhelming number of electrons will be substantially closer to the barrier than a factor of $e$, it seems likely that the average electron will have to travel a significantly shorter distance than the estimated mean free path before it falls below the barrier due to collisions. Thus, this naive estimate is likely not relevant for evaporation considerations. Without a more sophisticated calculation, it is not intermediately obvious what the relevant mean free path is and thus it is unclear whether this underlying assumption is indeed met. We provide such a calculation below.

Secondly, it is not clear whether the assumed functional form in Eq (3.) accounts for all collisions appropriately. The function only includes evaporation contributions from electrons with kinetic energy right at the barrier, and ignore the contributions from large angle collisions for higher-energy electrons not right at the barrier. It is unclear whether the contributions from these electrons are negligible for a typical set of experimental conditions. Furthermore, the assumed evaporation function assumes an average velocity slowing. Considering that the evaporation rate is the sum of discrete electron escapes, it is unclear whether utilizing an average slowing rate is appropriate.

\section{Model Overview}

In principle, a full molecular dynamics model of UCP electrons could be constructed to determine the electron evaporation rate in UCPs, but the necessary $O(N^{2})$ force calculations naively required make such a model computationally expensive. However, by assuming that the UCP electrons are in thermal equilibrium, approximations can be made that lead to drastically faster computational run times. Electrons in thermal equilibrium are, by definition, distributed in a Maxwell Boltzmann distribution. In the limit of $U>>k_bT$ only electrons in the upper tail of the Maxwellian distribution are able to escape from the confining potential. To take advantage of this fact, our simulation only tracks positions and velocities for electrons above a certain tracking threshold energy. Threshold energies were chosen to be low enough to not interfere with evaporation mechanics, but to be high enough to minimize computation time. A diagram illustrating the relationship between the tracking threshold and the overall electron distribution is given in Fig 1.

\begin{figure}
\includegraphics[width=3.5in]{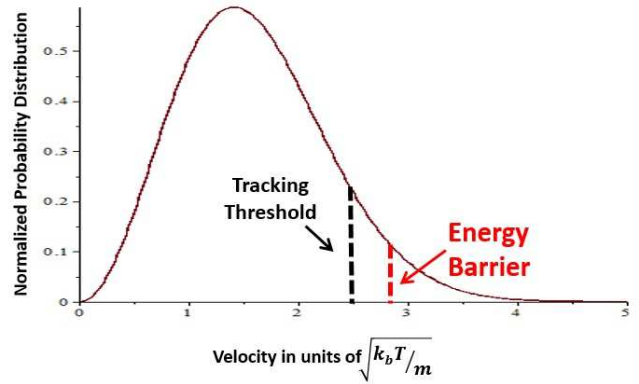}
\caption{A diagram illustrating the different velocity thresholds present in our model. The black dashed line corresponds to the tracking threshold. Electrons to the right of this line are tracked in the model, and electrons to the left are not tracked in the model. The red dashed line represents the energy barrier. Electrons to the right of this line are able to escape from the plasma and electrons to the left are not.}
\end{figure}

The thermal equilibrium assumption also allows for direct force calculations to be approximated by a series of random Coulomb collisions. Tracked electrons are assumed to be in contact with a reservoir of Maxwellian electrons, representing the complete electron distribution in the UCP. Tracked electrons collide with the reservoir electrons via a Monte Carlo collision operator, leading to changes in the momentum and energy of the tracked electron. In a similar fashion, tracked electrons also collide with a  reservoir of infinitely massive UCP ions, where treating the ions as having infinite mass is a reasonable approximation in UCPs. By utilizing such a method, an $O(N^2)$ process is reduced to an $O(N)$ process, greatly reducing computation time.

Occasionally, collisions cause tracked electrons to  lose sufficient energy so as to fall below the tacking threshold. When this occurs, these electrons are discarded from the simulation. To maintain detailed balance, a certain number of additional highly energetic electrons are generated at chosen time intervals. The process by which this was implemented will be discussed later in this section.

The tracked electrons in the model are uniformly distributed across the volume of a sphere, with the surface of the sphere acting as a ``hard wall'' potential barrier. When an electron comes in contact with the barrier, it will either escape if it has sufficient energy, or be reflected back toward the center if it does not. The rate at which electrons escape from the system, the evaporation rate, can thus be calculated.

In binary collisional theory, the probability of collision occurring in a time period, $dt$, is $n\sigma<v>dt$, where n is the particle density, $\sigma$ is the collisional cross section, and $<v>$ is the average difference velocity between a tracked electron and particles in either the electron or ion distribution. For electron-electron collisions, $<v>$ can be expressed as a function of tracked electron velocity, $v$:
\begin{equation}
<v>=\frac{k_bTerf\Big(v\sqrt{\frac{m}{2k_bT}}\Big)}{mv}+erf\Big(v\sqrt{\frac{m}{2k_bT}}\Big)v+\sqrt{\frac{2}{\pi}}e^{-\frac{mv^2}{2k_bT}}
\end{equation}
For electron-ion collisions, $<v>$ is set simply equal to v. 

For elastic collisions, when a collision occurs, the center of mass velocity of the two colliding particles is rotated by the angle $\chi$, changing the momentum of each particle\cite{Fund_PP}. The angle $\chi$ is defined by the following relationship:
\begin{equation}
\chi=2arctan(\frac{q_1q_2}{4\pi\epsilon_0\mu|\vec{v_1}-\vec{v_2}|^2b})
\end{equation}
 where $\mu$ is the reduced mass, b is the impact parameter, $q_1$ and $q_2$ are the charges of the two particles, and $\vec{v_1}$ and $\vec{v_2}$ are the 2 particle velocities.  Following the standard treatment for Coulomb collisions in a plasma, impact parameters are assumed to not exceed a maximum cutoff. For the purposes of this work, the standard $\lambda_D$ cutoff is assumed, where $\lambda_D$ is the Debye length.

Once a collision has been determined to have occurred, $b$ is randomly generated. Because ions in the model are assumed to be stationary, the deflection angle, $\chi$, for electron-ion collisions is solely a function of this generated impact parameter. However, for electron-electron collisions $\chi$ is also a function of the relative velocity of the two colliding electrons, $|\vec{v_1}-\vec{v_2}|$. The relative velocity, $|\vec{v_1}-\vec{v_2}|$, probability distribution as a function of $\vec{v_1}$, $\vec{v_2}$  is as follows:
\begin{equation}
|\vec{v_2}-\vec{v_1}|e^{-\frac{m\vec{v_1}\cdot \vec{v_1}}{2kT}}e^{-\frac{m\vec{v_2}\cdot \vec{v_2}}{2kT}}d^3\vec{v_1}d^3\vec{v_2}
\end{equation}
Since, in the context of the model, $\vec{v_1}$ is known, Eq. (9) can be reduced to exclusively a $\vec{v_2}$  probability distribution:
\begin{equation}
|\vec{v_2}-\vec{v_1}|e^{-\frac{m\vec{v_2}\cdot \vec{v_2}}{2kT}}d^3\vec{v_2}
\end{equation}
By randomly generating a value for $\vec{v_2}$, in addition to an impact parameter, the deflection angle, $\chi$, is fully determined.

As mentioned previously, modeling these collisions will result in electrons' energies falling below the tracking energy threshold.  Relatively quickly, this would mean that no electrons above the tracking energy would remain.  In steady-state, the fraction of tracked electrons should be approximately constant.  Thus, there needs to be some mechanism for ``creating'' electrons above the tracking threshold on a regular basis to maintain a Maxwellian distribution in the absence of electron evaporation. Care needed to be taken to generate a proper Maxwellian distribution. We elected to find a distribution of discrete velocities, henceforth known as a production function, that would lead to a Maxwellian steady state distribution in model simulations.

The process for developing a production function was as follows. First, a number of simulations were run where electrons were not allowed to escape. Electrons were added into the system at regular intervals with random positions and random velocity directions, but a fixed velocity magnitude, $v_i$. As the simulation ran, some electrons were removed as they went below the threshold velocity. Eventually, the rate of electrons being added and removed balanced out and the system reached a steady state number of electrons, resulting in the $i$th simulation having the electron velocity distribution, $f_i(v)$ for tracked electrons. The resulting velocity distributions from all of the simulations were then fit to a Maxwellian distribution through the following sum:

\begin{equation}
f(v)=A\sum_i a_if_i(v)
\end{equation}
where f(v) is the appropriate Maxwellian distribution, the $a_i$'s are normalized fit coefficients, and $A$ is a proportionality constant. Once the values of the $a_i$'s were found, the production function, $F(v)$ could be described: 

\begin{equation}
F(v)=\sum_ia_i\delta(v-v_i)dv
\end{equation}
Where the  best fit $a_i$ values represent the probability that an electron with the $i$th velocity magnitude will be generated. Once these values were determined, the production function could be utilized in simulations in which evaporation was included. 

\section{Results}

A number of model simulations were run to determine the impact of different plasma parameters on the electron evaporation rate, $\dot{N}$. Specifically, evaporation rates were calculated as a function of plasma spatial size, plasma depth, and electron temperature. For all simulations, ion and electron densities were both held constant at $1.35\times10^{13} m^{-3}$, corresponding to a set of low-density UCP experimental parameters. Timesteps were 10 ns long, and each simulation lasted 300 time steps. Evaporation rates were extracted from averaging the number of escapes occurring in the last 200 timesteps, during which the plasma was in a steady state.

Simulation results were used to test the veracity of scaling rules presented above. For instance, if $T$ is held constant, Eq. (6) can be expressed as a functional form:

\begin{equation}
\dot{N}=-AW^{\alpha}e^{-W}
\end{equation}
where $A$ and $\alpha$ are constants. The additional dependence on $W$ in Eq. (13) is typically present in expression for evaporation in ultracold atomic gases. To see if this form was reasonable, $\alpha$ and $A$ can be treated as a fit parameters. From Eq. (6), $\alpha$ would be expected to be 0. 

To test whether Eq. (13) expresses the proper functional form, evaporation rates were calculated for a series of different $W$ values at a constant $T$, and the resultant curve was fit to the functional form in Eq. (13). This process was then repeated for a series of different electron temperatures. The resultant $\alpha$ parameters from these fits can be seen in Table I:

\begin{table}
\resizebox{3.4in}{!}{
\begin{tabular}{|c|c|c|c|c|c|}
\hline
T&2K&3K&5K&7K&10K\\
\hline
$\alpha$&0.629&0.514&0.294&0.290&0.141\\
\hline
\end{tabular}
}
\caption{Table of the resultant best fit parameters from Eq. (13) from a fit to model calculations. Each column in the table represents a series of model calculations of the evaporation rate as a function of $W$ for a given constant $T$. The table gives the best fit parameter, $\alpha$, that corresponds to each of the constant $T$ curves.} 
\end{table}

These results show that $\alpha$ is consistently greater than 0, indicating that evaporation is less strongly dependent on $W$ than Eq.(6) would predict. Additionally, since $\alpha$ increases with decreasing temperature, evaporation becomes more weakly dependent on $W$ as $T$ decreases. These results show that the functional form of the electron evaporation rate from Eq. (13) is itself incorrect, and thus evaporation cannot be modeled in such a manner. This point is further illustrated in Fig 2.

\begin{figure}
\includegraphics[width=3.5in]{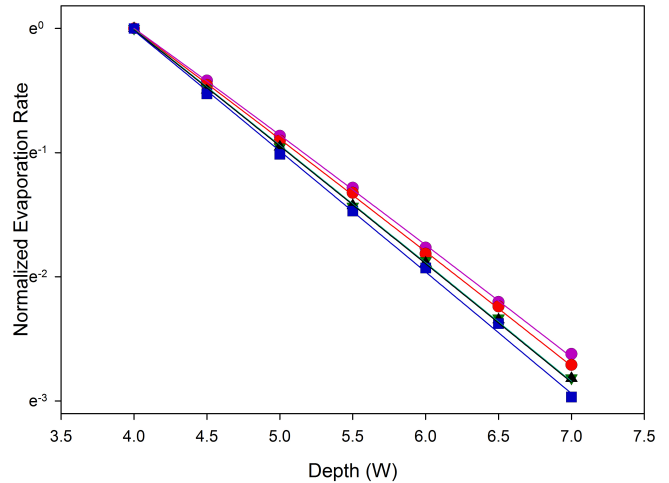}
\caption{Plot of the resultant best fit curves from Eq. (13) being fit to model calculations. Each curve is fit to a series of model calculations which vary in $W$ but are held at a constant $T$. The purple points correspond to a $T$=2K curve, the red points are $T$=3K, the green points are $T$=5K, the black points are $T$=7K, and the blue points are $T$=10K. The data points are model calculations of $\dot{N}(W)/\dot{N}(W=4)$, and fit curves are represented by the corresponding curves shown. The plot shows that the curves scale differently. This suggests that Eq. (13) does not accurately describe $\dot{N}$. Note that the y-axis is on a log scale.}
\end{figure}

Simulation results were also used to test the temperature dependence of evaporation at constant depth to temperature ratio (i.e. constant $W$). At a constant $W$, Eq. (6) reduces to the following functional form: 

\begin{equation}
\dot{N}=k_1T^{-k_2}ln(-k_3T^{\frac{3}{2}})
\end{equation}

where $k_1$,$k_2$, and $k_3$ are constants. Since we are only testing scaling laws at this point, these constants can be treated as fitting parameters. By using a method analogous to to the constant $T$ case, the $T^{-\frac{3}{2}}$ scaling, implied by Eqs. (6) and (14), can again be tested. This leads to the results shown in Table II:

\begin{table}
\resizebox{3.4in}{!}{
\begin{tabular}{|c|c|c|c|c|c|c|c|}
\hline
$W$&4&4.5&5&5.5&6&6.5&7\\
\hline
$k_1$&0.522&0.512&0.538&0.484&0.499&0.505&0.590\\
\hline
$k_2$&0.866&0.911&0.943&0.941&0.932&0.947&1.055\\
\hline
$k_3$&4.041&4.823&4.395&6.646&5.653&5.690&4.130\\
\hline
\end{tabular}
}
\caption{Table of the resultant best fit parameters from Eq. (13) being fit to model calculations. Each $W$ in the table represents a series of model calculations which vary in $T$ but are held at a constant $W$, and $k_1$ $k_2$ and $k_3$ are the best fit parameters corresponding to this constant $W$ curve.} 
\end{table}

These results are inconsistent with a $T^{-\frac{3}{2}}$ scaling given the value of $k_2$. Additionally, Eq. (14) suggests that $k_2$ and $k_3$ should stay constant as a function of $W$, which is also contradicted by the our calculated results. These results show that the functional form of the electron evaporation rate from Eq. (14) is incorrect for the given experimental conditions. Examples of these fits can be seen in Fig 3.

\begin{figure}
\includegraphics[width=3.5in]{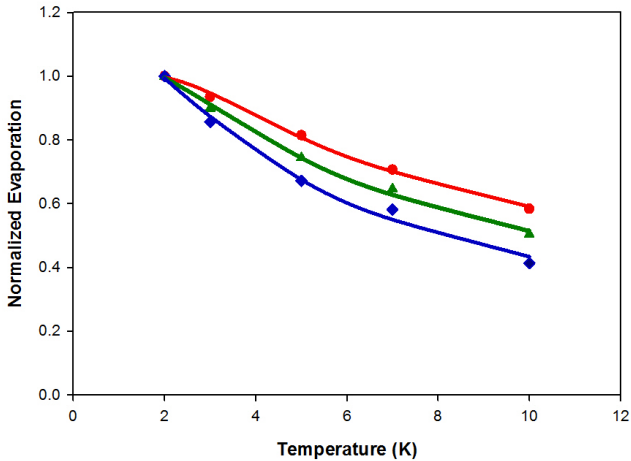}
\caption{Plot of the resultant best fit curves from Eq. (14) being fit to model calculations. Each curve is fit to a series of model calculations which vary in $T$ but are held at a constant $W$. The red points correspond to a $W$=4 curve, the green points correspond to a $W$=5 curve, and the blue points correspond to a $W$=7 curve. The data points represent model calculation of $\dot{N}/\dot{N}(T=2)$, and fit curves are represented by the corresponding curve. The plot shows that not all of the curves scale identically. This suggests that Eq. (14) does not accurately describe $\dot{N}$.}
\end{figure}

The failure of Eq. (6) to properly predict the proper parameter scaling of the evaporation rate suggests that at least one of the two underlying assumptions of the theory is incorrect for typical UCP experimental conditions. The first of these assumptions states that when a collision excites an electron above the barrier, it is immediately considered to have escaped. However, if the plasma size is larger than the effective mean free path, it becomes likely that a secondary collision de-excites that same electron back below the barrier, preventing the electron from escaping. The only exception is if the electron is at the edge of the UCP. Since it is unclear what the effective mean free path is in the context of evaporation, it is questionable whether or not this assumption is valid for typical UCP collisions.

To characterize the impact that the mean free path will have on evaporation, we found it useful to consider two limiting cases. In the case where the mean free path is much larger than the size of the plasma, the absolute evaporation should scale linearly with the electron number without any dependence on the plasma radius, $R$, assuming a constant electron density. Conversely, if the plasma is much larger than the mean free path, electrons in close proximity to the plasma edge should contribute much more heavily to the evaporation rate than other electrons. Presumably, in the limit of an infinitely large UCP, evaporation should scale with UCP surface area. However, if the electron density is held constant, the electron number scales with the UCP volume, which leads to the evaporation rate per electron, $\dot{N}/N$, scaling as $1/R$ in this surface-dominated limit.

To determine where plasmas with the parameters used in the simulation were between these two limits, we investigated how $\dot{N}/N$ varied with plasma spatial size, $R$. Simulations were run varying the UCP size and electron number in a manner consistent with a constant density, and a per electron evaporation rate vs UCP size curve was generated. The results of these calculations, for one set of conditions, can be seen in Fig 4.

\begin{figure}
\includegraphics[width=3.5in]{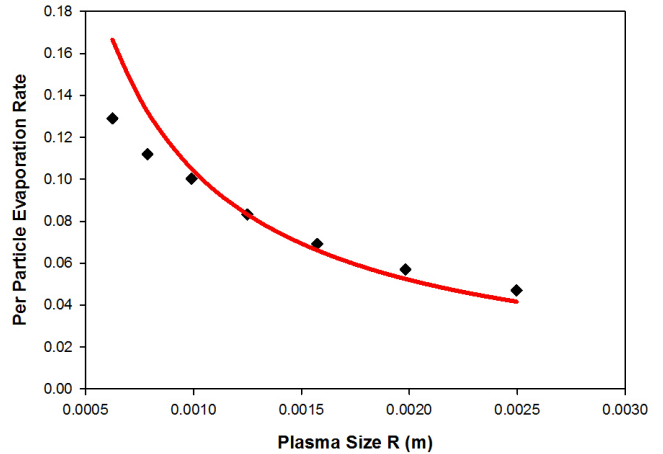}
\caption{Plot of $\dot{N}/N$ as a function of plasma size, $R$, for a T=10K and W=4 plasma. The black data points are the model calculations, and the red line is a $1/R$ curve. For larger values of $R$, model calculations scale roughly as $1/R$ indicating that the plasma size is much greater than the typical mean free path for these conditions.}
\end{figure}

The figure shows that per particle evaporation scaled roughly as $1/R$ at larger values of $R$, which is consistent with the typical mean free path being much smaller than the plasma spatial size. At smaller plasma sizes, this scaling became shallower. Constant evaporation scaling with $R$, the scaling implicitly assumed in Eq. (6), was not observed over the investigated range of parameters. These results show that the underlying assumption in Eq (6) about the mean free path is incorrect. Furthermore, for typical simulation plasma parameters, the evaporation rate is consistent with mean free path effects being dominant. It is therefore desirable to quantify the magnitude of these effects. 

To do this, we introduce the concept of a effective electron evaporation source density, $n_{eff}$. In general, electrons that are closer to the edge are more likely to escape the plasma. The purpose of the effective density is to weight these more-likely-to-leave electrons more highly than their counterparts near the plasma center, and to quantify how this weighting changed as a function of $R$. To do this we utilized the following simple approximate model:

\begin{equation}
n_{eff}=ne^{-\frac{R-r}{c}}
\end{equation}
where $n$ is the electron density, $R$ is the plasma size, $r$ is the standard radial coordinate, and $c$ is an effective evaporation skin depth quantifying the degree of which mean free path effects impact evaporation. Utilizing this approximate description, we assume that $\dot{N}$ takes the following form:

\begin{equation}
\dot{N}=-\kappa N_{eff}=-4\pi \kappa \int_{0}^{R}r^2 n_{eff}dr
\end{equation}
 where $\kappa$ is a proportionality constant, and $N_{eff}$ is the effective number derived from $n_{eff}$. Combining Eq. (15) and Eq. (16) and evaluating the integral yields the following expression:

\begin{equation}
\dot{N}=-4\pi n\kappa (R^2-2Rc+2c^2-2c^2e^{-R/c})
\end{equation}
By treating $\kappa$ and $c$ as fit parameters, Eq. (17) was fit to results similar to those in Fig. 4, and an effective evaporation skin depth was extracted.

\begin{figure}
\includegraphics[width=3.5in]{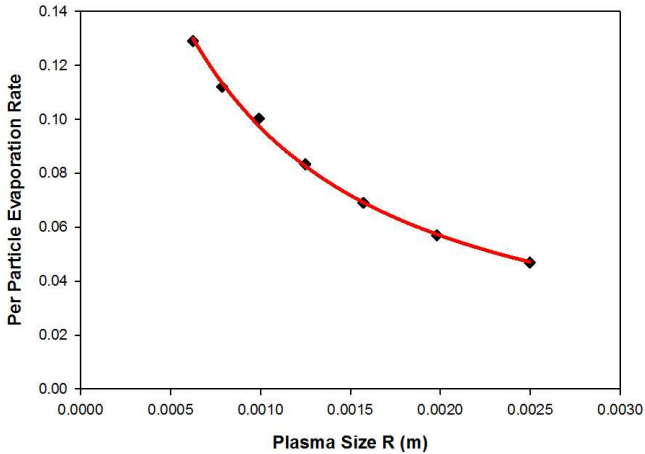}
\caption{The plot shows Eq. (17) being fit to model calculations of $\dot{N}/N$ as a function of $R$. The black data points are the model calculations for a T=10K and W=4 plasma, and the red line is the resultant fit function. The plot shows that Eq. (17) describes the model results well.}
\end{figure}

The resultant fit parameters, $\kappa$ and $c$, can give insight about the applicability of the assumed analytical evaporation functional form, even in the absence of mean free path considerations. In the limit of $R<<c$, the model predicts an evaporation rate of $\kappa$. If the functional form of Eq.(6) is correct, its predicted evaporation rate should match our calculated $\kappa$. This was not the case, however, as we observed $\kappa$ to be 3-7 times smaller than would be implied by Eq. (6). This was, at least in part, due to the electron velocity distribution.  Eq. (6) assumes a Maxwellian, but a collection of electrons in a potential well will not form a Maxwellian distribution in steady state. At the low electron energies, these distributions are roughly the same, but the Maxwellian will significantly overestimate the electrons near the barrier. This leads to an overestimation of $dN/dt$ in Eq. (6), and subsequently an overestimation in the evaporation rate.

\begin{figure}
\includegraphics[width=3.5in]{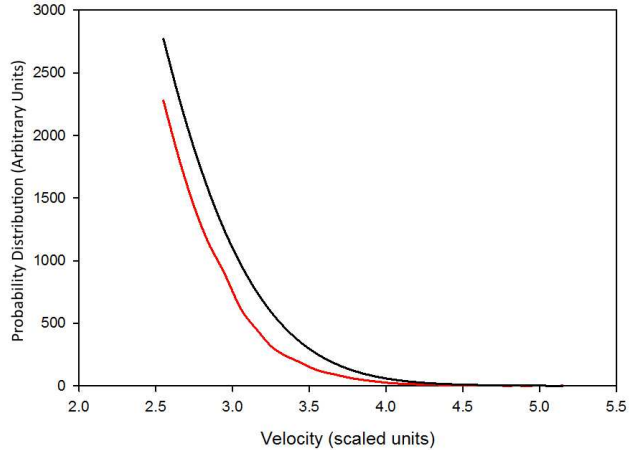}
\caption{The plot shows the discrepancy between a Maxwellian velocity distribution and the equilibrium velocity distribution for electrons in a finite potential well. The black curve represents an ideal Maxwellian, and the red curve is the model calculated velocity distribution for a $W$=5, $T$=10K plasma.} 
\end{figure}

The mean free path fit parameters were also used to test of the scaling of the $dv/dt$ component of the evaporation functional form in the limit of $R<<c$. We compared our calculated values of $c$ to an estimate involving the form of $dv/dt$ in Eq. (5) above. In the limit of small velocity changes, for a given initial velocity, $v_0$, above the barrier, the characteristic time,$\tau$, it takes to decay to the barrier velocity, $v_U$, can be defined:

\begin{equation}
\tau(v_0)=\frac{v_0-v_U}{(dv/dt)|_{v_0}}
\end{equation}

This implies an approximate velocity dependent mean free path of $c(v)=v\tau(v)$. By integrating over the velocities of all of the electrons above the barrier an average $c$ can be calculated. For typical experimental conditions, such a calculation resulted in a $c$ on the order of about a mm, an order of magnitude larger than suggested by  the results from the prior mean free path fit. The results of this calculation suggests that the average electron velocity slowing, $dv/dt$, is not the relevant rate with regards to evaporation. This is likely due to $dv/dt$ being an average quantity. The evaporation rate is the sum of discrete electron escape events, and it is not immediately obvious that an average rate would properly account for the relevant physics. In addition, the average escape path for model electrons  can often be much longer than $R$. A typical escaping electron will undergo a number of deflecting collisions over the course of its escape, which presumably effectively lengthen the electron's escape path.

Unfortunately, the previously developed analytical expressions seem to not be applicable to ultracold plasmas, at least under the conditions studied. Eq. (6) incorrectly predicted electron evaporation rates and did not scale correctly. While analytical expressions that accurately predict the electron evaporation rate could presumably be developed, such expressions do not currently exist. Thus, a numerical model, such as the one developed in this work, will be needed to properly calculate the electron evaporation rate in ultracold plasma systems.

\section{Conclusion}
We have developed a model that calculates the rate evaporation from an ultracold plasma. Model results were compared to previously developed analytical expressions for evaporation. These expressions proved inconsistent with our results as the model scaled with plasma parameters differently than the simple evaporation expressions predicted. Furthermore, we demonstrated that this discrepancy can, at least partially, be explained by the finite size of the plasmas examined in this work, and that absence of size considerations is a limitation of these previously developed expressions. This work demonstrates that such simple scaling rules are not accurate when used to calculate evaporation in UCPs, and that a model like the one developed in this work is needed.
\section{Acknowledgments}
We acknowledge support of the Air Force Office of Scientific Research (AFOSR), grant number FA9550-12-1-0222.

\bibliography{UCP}
\end{document}